\documentclass[letterpaper, 10 pt, conference]{ieeeconf}   
\IEEEoverridecommandlockouts                               
\usepackage{optidef}

\usepackage{amssymb,amsmath,color}
\usepackage{graphicx}
\usepackage{xspace,stackrel,hyperref}
\usepackage{mathtools}
\usepackage{tikz,pgfplots}
\usepackage{autonum}
\usepackage{dsfont}
\usepackage{etoolbox}

\usepackage{cite}

\usepackage{arydshln}

\usepackage{esvect}
\usepackage{optidef}
\usepackage{multirow}

\usepackage[ruled,vlined]{algorithm2e}
\usepackage{lipsum,adjustbox}

\DeclareMathOperator*{\argmax}{arg\,max}
\DeclareMathOperator*{\argmin}{arg\,min}

\definecolor{mycolor4}{RGB}{230,97,1}
\definecolor{mycolor2}{RGB}{178,171,210}
\definecolor{mycolor3}{RGB}{253,184,99}
\definecolor{mycolor1}{RGB}{94,60,153}

\makeatletter 
\pretocmd\@bibitem{\color{black}\csname keycolor#1\endcsname}{}{\fail}
\newcommand\citecolor[1]{\@namedef{keycolor#1}{\color{blue}}}
\makeatother

\DeclareMathAlphabet{\mathcal}{OMS}{cmsy}{m}{n}

\def\beq{\begin{equation}}
\def\eeq{\end{equation}}

\newcommand{\mc}{\mathcal}

\newcommand{\Z}{\mathbb{Z}}

\newcommand{\R}{\mathds{R}}

\newcommand{\defineas}{\coloneqq}

\newcommand{\norm}[1]{\left\lVert#1\right\rVert}

\definecolor{mycolor1}{RGB}{230,97,1}
\definecolor{mycolor2}{RGB}{178,171,210}
\definecolor{mycolor3}{RGB}{253,184,99}
\definecolor{mycolor4}{RGB}{94,60,153}
\definecolor{mycolor5}{rgb}{0,0,0}

\tikzset{
  pics/car/.style args={#1}{
     code={
     \begin{scope}[scale=0.15]
      \shade[top color=#1, bottom color=white, shading angle={135}]
        [draw=black,fill=red!20,rounded corners=0.2ex] (1.5,.5) -- ++(0,1) -- ++(1,0.3) --  ++(3,0) -- ++(1,0) -- ++(0,-1.3) -- (1.5,.5) -- cycle;
    \draw[ rounded corners=0.5ex,fill=black!20!blue!20!white]  (2.5,1.8) -- ++(1,0.7) -- ++(1.6,0) -- ++(0.6,-0.7) -- (2.5,1.8);
    \draw[thick]  (4.2,1.8) -- (4.2,2.5);
    \draw[draw=black,fill=gray!50,thick] (2.75,.5) circle (.5);
    \draw[draw=black,fill=gray!50,thick] (5.5,.5) circle (.5);
    \end{scope}
     }
  }
}

\newtheorem{assumption}{Assumption}

\newtheorem{theorem}{Theorem}
\newtheorem{proposition}{Proposition}
\newtheorem{corollary}{Corollary}
\newtheorem{definition}{Definition}

\newtheorem{remark}{Remark}

\pgfplotsset{compat=1.17}
\usetikzlibrary{pgfplots.groupplots}

\newtoggle{full_version}
\toggletrue{full_version}

\title{\LARGE \bf
Learning How to Price Charging in Electric Ride-Hailing Markets
}

\author{Marko Maljkovic, Gustav Nilsson, and Nikolas Geroliminis
\thanks{M.~Maljkovic, G.~Nilsson, and N.~Geroliminis are with the School of Architecture, Civil and Environmental Engineering, École Polytechnique Fédérale de Lausanne (EPFL), 1015 Lausanne, Switzerland. {\tt\small \{marko.maljkovic, gustav.nilsson, nikolas.geroliminis\}@epfl.ch}.}%
\thanks{This work was supported by the Swiss National Science Foundation under NCCR Automation, grant agreement 51NF40\_180545.}
\iftoggle{full_version}{}{\thanks{An extended version containing all the proofs is available at \url{http://arxiv.org/abs/2203.09327}}}
}

\begin{document}

\maketitle
\thispagestyle{empty}
\pagestyle{empty}

\begin{abstract}
With the electrification of ride-hailing fleets, there will be a need to incentivize where and when the ride-hailing vehicles should charge. In this work, we assume that a central authority wants to control the distribution of the vehicles and can do so by selecting charging prices. Since there will likely be more than one ride-hailing company in the market, we model the problem as a single-leader multiple-follower Stackelberg game. The followers, i.e., the companies, compete about the charging resources under given prices provided by the leader. We present a learning algorithm based on the concept of contextual bandits that allows the central authority to find an efficient pricing strategy. We also show how the exploratory phase of the learning can be improved if the leader has some partial knowledge about the companies’ objective functions. The efficiency of the proposed algorithm is demonstrated in a simulated case study for the city of Shenzhen, China.

\end{abstract}

\section{Introduction}

With the widespread adoption of electric vehicles (EVs) as an eco-friendly mode of transportation, the need for reliable and efficient charging infrastructure has emerged as a crucial factor dictating their overall usability~\cite{9318522}. To profitably manage large electric fleets in the near future, big ride-hailing companies such as Uber, Lyft, etc., would likely have to devise intelligent charging strategies dictated by the spatio-temporal distribution of the power supply. On the other hand, due to the ever-increasing electricity demand, coordinated charging of large fleets could have a strong positive impact on preventing imbalances and overloads in the power network. From the perspective of the local authorities, the interplay between the energy stakeholders, the ride-hailing service providers, and the heterogeneous demand opens the door for trading different services to achieve a societal optimum in terms of energy management and congestion levels in the region. As the company operators strive to reduce their operational expenses, which among others include charging costs, there is an inherent need to minimize queuing at charging stations due to the limited capacity of the shared infrastructure. In return, this creates a competitive environment between companies as stations in high-demand areas could be more prone to overcrowding. 

\begin{figure}
    \centering
    \resizebox{.3\textwidth}{!}{%
    \begin{tikzpicture}[scale=0.8]

        \draw[top color= white, bottom color=white, draw=black] (-2.5, -4.5) rectangle (0.0, -5.5);
        \node (f2) at (-0.55,-5.0){$\mc C_2$};
        
        \draw[top color= white, bottom color=white, draw=black] (-5.1, -4.5) rectangle (-2.6, -5.5);
        \node (f1) at (-3.15,-5.0){$\mc C_1$};
        
        \draw[top color= white, bottom color=white, draw=black] (2.6, -4.5) rectangle (5.1, -5.5);
        \node (fN) at (4.55,-5.0){$\mc C_{N}$};
        
        \node (myfirstpic1) at (-1.75,-5.0) {\includegraphics[width=.05\textwidth]{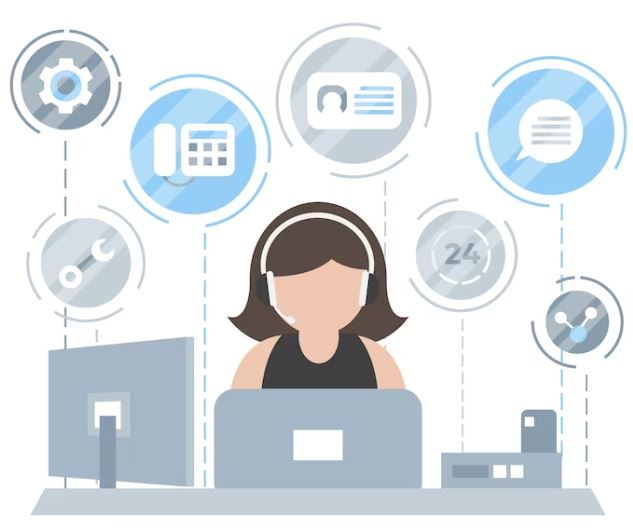}};
        \node (myfirstpic2) at (-4.35,-5.0) {\includegraphics[width=.05\textwidth]{figures/operator.png}};
        \node (myfirstpic3) at (3.35,-5.0) {\includegraphics[width=.05\textwidth]{figures/operator.png}};

        \node (d1) at (1.3, -5.0)[circle,fill,inner sep=0.75pt]{};
        \node (d2) at (1.5, -5.0)[circle,fill,inner sep=0.75pt]{};  
        \node (d3) at (1.1, -5.0)[circle,fill,inner sep=0.75pt]{};   
        
        \draw[dashed] (-5.2, -4.3) rectangle (5.2, -5.7);
        
        \draw[dashed] (-5.4, 0.7) rectangle (5.4, -5.9);
        
        \draw[->] (0.0, -3) -- (0.0, -4.3);
        
        \draw[top color= white, bottom color=white, draw=black] (-2.5, 0.0) rectangle (2.5, -3);
        
        \node (myfirstpic4) at (-1.7,-1.5) {\includegraphics[width=.07\textwidth]{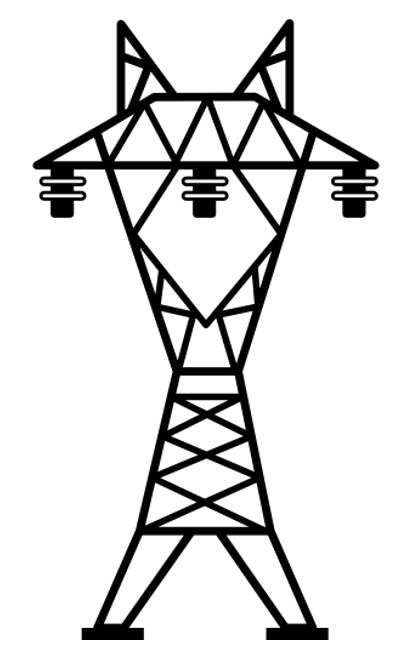}};

        \node (myfirstpic5) at (0.8,-1.5)
        {\includegraphics[width=.15\textwidth]{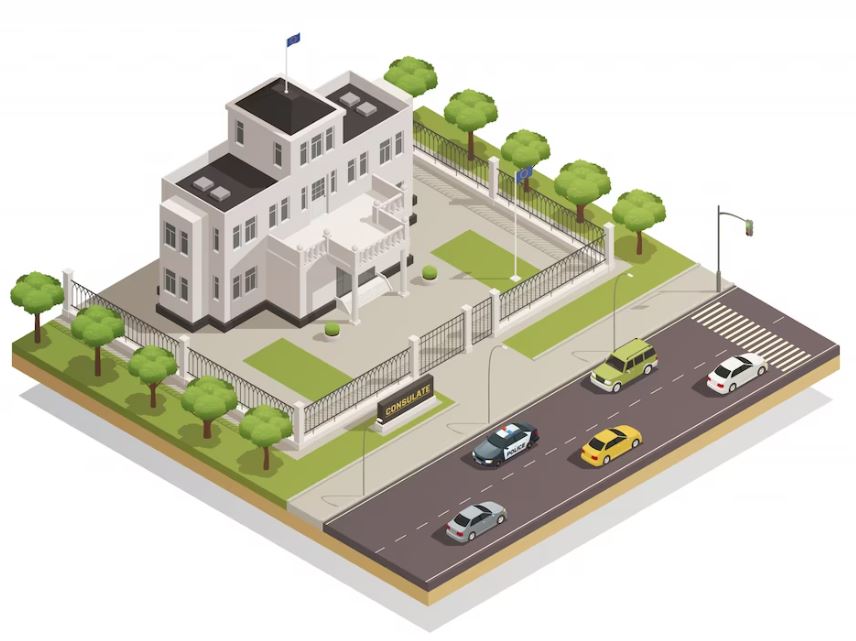}};

        \node[] (D5) at (0.0, 0.3){Central authority};
        
        \node[] (D3) at (0.3, -3.65){$\pi_t$};
        
        \node[] (G0) at (-3.6, -4.0){Market $\mc M\left(\mc C;\pi_t\right)$};
        \node[] (G1) at (-3.5, 1.0){Stackelberg game $G_L$};
        
    \end{tikzpicture}%
    }

    \caption{Schematic sketch of the bi-level problem setting. The $|\mc C|=N$ ride-hailing companies form the market $\mc M(\mc C)$, parametrized by the vector of charging prices determined by the central authority.}
    \label{fig:sketch}
\end{figure}

We envision that the central authority acts as a regional entity of sufficient regulative power. As illustrated in Figure~\ref{fig:sketch}, we assume the central authority is in charge of determining the charging price at each charging station. Regardless of whether the central authority is the government, the power grid operator, etc., we assume it aims to steer the outcome of the competition between the ride-hailing companies towards the system optimum by offering discounted charging at certain stations. Consequently, through smartly designed prices, the central authority would hope to motivate the management of the ride-hailing companies to also charge their fleets in the more distant areas. As a result, the central authority would reduce the burden both on the power grid and the traffic network in the demand-attractive areas. In any case, with such a pricing-oriented structure, the interactions between the central authority and the ride-hailing market align well with the concept of Stackelberg games.

We aim to extend the analysis of the bi-level pricing game previously introduced in~\cite{9838005,Hierarchical,ECC2023}, where the system optimum is introduced by a predefined vehicle distribution profile. By adopting the operational cost model based on~\cite{9102356, article4, 9541309}, the ride-hailing market becomes governed by a quadratic aggregative game. Upon the announcement of the charging prices, we assume that the rational company operators choose a no-regret strategy given by the Nash equilibrium (NE). Previous studies~\cite{9838005,Hierarchical,ECC2023} show what can be achieved by means of collaborative information exchange between the agents on different hierarchy levels. Here, we analyze the pricing problem from the perspective of the leader and start by treating the ride-hailing market as a 'grey box' environment that provides little knowledge about the lower-level game. To begin with, we assume the central authority has no information about the functional form of the costs and constraints governing the operation of the companies. For such a scenario, we propose a contextual bandit (CB) algorithm~\cite{9185782} that learns how to price charging based on some intermediate aggregate information about the charging demand of the companies. In essence, the environment encapsulates the pricing-induced interactions within the ride-hailing market and outputs just the attained NE. The algorithm then trains the parameters of the pricing module by observing the pairs of applied prices and observed NE. The problem of learning Stackelberg equilibria, under certain assumptions on the game structure, has already been studied in the literature~\cite{pmlr-v119-fiez20a}. In the context of pricing, there is a large body of research applying the multi-step counterpart of the contextual bandits, i.e., the reinforcement learning (RL) algorithm~\cite{9631970,9536423,9210540}. However, the setup of our problem does not facilitate a Markov decision process (MDP) based state transition as the prices applied at a particular instant of time have no influence on the future state of the environment described mainly by the charging demand of the vehicles. As such, our setup is not compatible with the RL-based methods. 

To the best of our knowledge, there are no works
directly applicable to our problem of learning the central authority's pricing strategy in a Stackelberg game with limited information about the lower-level ride-hailing market. Naturally, we complement the proposed CB-based algorithm with an analysis of the pricing procedure with respect to different levels of information available to the leader. Namely, if the central authority has access to the cost functions but not to the constraints governing the ride-hailing market, we propose how to construct an initial exploration space for gathering high-quality training data when there is no previous domain-based knowledge. Finally, if the central authority also has information about the constraints in the ride-hailing market, we show that there is no need to formulate a learning problem since the bi-level Stackelberg pricing game reduces to a mathematical program with complementarity constraints (MPCC)~\cite{10.2307} that can be recast into an instance of a mixed integer linear or quadratic problem (MILP/MIQP)~\cite{KLEINERT2021100007} using the big-M reformulation~\cite{bigm}.          

The paper is outlined as follows: the rest of this section is devoted to introducing the basic notation. In Section~\ref{sec:problemsetup}, we introduce the general problem setup before defining the structure of the electric ride-hailing market in Section~\ref{sec:model}. In Section~\ref{sec:alg}, we then present our main methodological and theoretical results. Finally, we conclude the paper with Sections~\ref{sec:example} and~\ref{sec:conclusion}, where we test our method in a numerical case study and propose ideas for future research.

\textit{Notation:} Let $\R$ denote the set of real numbers, $\R_+$ the set of non-negative reals, and $\Z_{>0}$ the set of positive integers. Let $\mathbf{0}_{m}$ and $\mathbf{1}_{m}$ denote the all zero and all one vectors of length $m$ respectively, and $\mathbb{I}_{m}$ the identity matrix of size $m \times m$. For a finite set $\mc A$, we let $\R_{(+)}^{\mc A}$ denote the set of (non-negative) real vectors indexed by the elements of $\mc A$ and $\left|\mc A\right|$ the cardinality of $\mc A$. Furthermore, for finite sets $\mc A$, $\mc B$ and a set of $|\mc B|$ vectors $x^i\in \R_{(+)}^{\mc A}$, we define $x \defineas \text{col}\left((x^{i})_{i\in \mc B}\right)\in \R^{|\mc A||\mc B|}$ to be their concatenation. For $A\in \R^{n\times n}$, $A\succ 0 (\succeq 0)$ is equivalent to $x^TAx>0 (\geq 0)$ for all $x\in\R^{n\times n}$. We let $A\otimes B$ denote the Kronecker product between two matrices and for a vector $x\in\R^n$, we let $\text{diag}(x)\in\R^{n\times n}$ denote a diagonal matrix whose elements on the diagonal correspond to vector $x$. For matrices $M_i$, such that $i\in\mc A$, we let $\text{Diag}(M_i)_{i\in\mc A}$ denote their concatenation into a block-diagonal matrix. For a matrix $P$, let $\textbf{\text{tr}}(P)$ denote its trace.

\section{Problem statement}\label{sec:problemsetup}
We consider a Stackelberg pricing game where the central authority, denoted as the leading agent $L$, wants to steer the decisions made by the agents in the ride-hailing market $\mc M(\mc C)$. The market is defined by the operational management of a set of ride-hailing companies $i\in\mc C$ operating in a region where access to a set of shared charging stations $\mc H$ for electric vehicles is offered, with $|\mc C|=N$ and $|\mc H|=M$. We analyze the problem from the short-term perspective, i.e., for one snapshot of the day in which multiple drivers from the ride-hailing companies would like to recharge. We assume that the operator of each company is responsible for coordinating the charging of the respective electric fleet, and aims to do so in an attempt to minimize the one-step-ahead operational cost. Namely, each operator aims to match the vehicles that want to recharge with the stations so as to optimize a cost that encompasses the cost of charging, the monetary equivalent of the time spent queuing at the charging stations, and the expected revenue induced by a particular coordinated charging strategy. The central authority, on the other hand, parametrizes the optimization problem of each company by choosing the charging prices $\pi\in\mc P\subseteq\R^{M}$ for the stations in the region. For each $i\in\mc C$, let $N_i$ be the number of vehicles that want to recharge and $n=\text{col}((N_i)_{i\in\mc C})$. Then, the decision variable of company $i\in\mc C$ is given by $x^i\in\mc X_i\subseteq\R^M$, with $\lVert x^i\rVert_1=N_i$ and $x^i_j\geq 0$ representing the number of vehicles assigned to station $j$. We let the sets $\mc X_i$ encode the constraints of each company. Since $x^i$ is chosen as a real vector, and the number of vehicles that can be sent to each station is an integer, for a perfect match to exist between the vehicles and the stations, it suffices to choose polytopic constraints as previously discussed in~\cite{9838005}. Therefore, we assume the sets $\mc X_i$ are given by: 
\begin{equation}
    \mc X_i\defineas\left\{x^i\in\R^{M}\mid A_ix^i=b_i \land G_ix^i\leq h_i\right\}\,,
    \label{eq:constrset}
\end{equation}
with $A_i=\textbf{1}^T_{M}$, $b_i=N_i$, $G_i\in\R^{m_i^{\text{ineq}}\times M}$ and $h_i\in\R^{m_i^{\text{ineq}}}$, for properly chosen $m_i^{\text{ineq}}\in\Z_{>0}$. If we define sets $\mc X\defineas \prod_{i\in\mc C} \mc X_{i}$ and $\mc X_{-i}\defineas \prod_{j\in\mc C\setminus i} \mc X_{j}$, then the joint strategy of all followers can be denoted as $x \defineas \text{col}\left((x^{i})_{i\in \mc C}\right)\in\mc X$ and for every $i\in\mc C$, we can define $x^{-i} \defineas \text{col}\left((x^{j})_{j\in \mc C \setminus i}\right)\in\mc X_{-i}$. The objective of each company is to minimize the cost 
\begin{equation}
    \label{eq:pricingcost}
    J^i\left(x^i, x^{-i} ; \pi\right)\defineas \hat{J}^i\left(x^i, x^{-i}\right)+\left(x^i\right)^TS_i\pi\,,
\end{equation}
where the first term $\hat{J}^i\left(x^i, x^{-i}\right)$ encapsulates the influence of other companies on the perceived cost, such as queuing at the station and lost income from passengers,  and the second term describes the total charging price to be paid. Here, the matrix $S_i\in\R^{M\times M}$ is diagonal, i.e.,  $S_i=\text{diag}\left(d^i\right)\succeq 0$, and every element $d^i_j\in\R_{+}$ of the vector $d^i\in\R^M_+$ can be interpreted as the expected average charging demand per one vehicle of the $i$-th company when choosing the station $j\in\mc H$. The $\pi$-parametetrized ride-hailing market can now be described as a set of $N$ optimization problems given by:
\begin{equation}
    \mc M\left(\mc C;\pi\right)\defineas \left\{\min_{x^{i}\in \mc X_{i}}J^{i}\left(x^{i},x^{-i};\pi\right),\forall i\in \mc C\right\} \,.
    \label{eq:market}
\end{equation}

We assume the companies in the market $\mc M\left(\mc C;\pi\right)$ collaborate to play a no-regret strategy according to a Nash equilibrium (NE) given in the following definition.
\begin{definition}\label{def:NE}
    For any $\pi\in\mc P$, a joint strategy $x^*\in\mc X$ is a NE of the game played in $\mc M\left(\mc C;\pi\right)$, if for all $i\in\mc C$ and all $x^i\in\mc X_i$ it holds that $J^i\left(x^{i*}, x^{-i*}; \pi\right)\leq J^i\left(x^{i}, x^{-i*};\pi\right)$
\end{definition}
\medskip
Specifically, we focus on a subset of NE given by Definition~\ref{def:NE}, known as the variational Nash equilibria (v-NE).
\begin{assumption} \label{ass:1}
    For any $\pi\in\mc P$, the companies in $\mc M\left(\mc C;\pi\right)$ play a joint v-NE $x\in\mc V_{\pi}(\mc M)$ described by the set $\mc V_{\pi}(\mc M)\defineas\{x\in\mc X|(y-x)^TF(x,\pi)\geq 0,\:\forall y\in\mc X\}$,
    where $F(x,\pi)\defineas \text{col}((\nabla_{x^{i}}J^{i}(x^i, x^{-i};\pi))_{i\in \mc C})$.
\end{assumption}
\medskip

In this paper, we assume the central authority is interested in controlling vehicle distribution among charging stations. The central authority chooses the prices $\pi\in\mc P$ in an attempt to force the company operators to coordinate charging such that the resulting total vehicle distribution matches a predefined one given by vector $\mc Z\in\left[0,1\right]^M$ with $\textbf{1}^T\mc Z=1$. 

For every $i\in\mc C$, let $\Lambda_i\in\R^{M\times NM}$ be a selection matrix
$\Lambda_i\defineas\left[\textbf{0}_{M\times (i-1)M}\:|\:\mathbb{I}_{M\times M}\:|\:\textbf{0}_{M\times (N-i)M}\right]$, $\Lambda\defineas\sum_{i\in\mc C}\Lambda_i$, $\Lambda_{-i}\defineas\Lambda-\Lambda_i$ and $\overline{\Lambda}\defineas\textbf{1}_{N-1}^T\otimes \mathbb{I}_M$. Then, the central authority's optimization problem can be cast as
\begin{equation}
        G_L:=\left\{\begin{array}{c}
        \displaystyle\min _{\pi \in\mc P}  J^{L}\left(x^{*},\pi\right)=\frac{1}{2}\norm{\Lambda x^*-\textbf{1}^Tn\mc Z}_2^2 \\
        \text { s.t. } 
        x^*\in\mc V_{\pi}\left(\mc M\right)
        \end{array}\right\} \,.
        \label{eq:SG}
\end{equation}

If $\mc K$ denotes the space of available partial observations of the market state, then the central authority would like to obtain a functional $\pi:\mc K\rightarrow\mc P$ that would map each observation into properly chosen charging prices at stations~$\mc H$. It is evident that the amount of shared information about the structure of the market, i.e., the cost functions $\hat{J}^i\left(x^i, x^{-i}\right)$ and constraint sets $\mc X_i$, directly dictates to what extent the problem~\eqref{eq:SG} can be analytically solved. In cases when knowledge about the model is scarce, the central authority would try to learn the market behavior and how to price better through interactions with the ride-hailing market.

In the following section, we will define the elements of the underlying operational cost model of ride-hailing companies. However, in Section~\ref{sec:alg}, we will start by treating this model as a black box and show that it is possible to design a general learning-based method capable of tackling such a market structure. We will then show how the initial exploration space~$\mc P$ for the learning-based method can be designed should the companies be willing to disclose the information about the costs $\hat{J}^i\left(x^i, x^{-i}\right)$ and how the complete learning procedure collapses to an instance of a mixed integer linear program (MILP) if $\mc X_i$ is available as well.   
\definecolor{m1}{RGB}{250,200,200}
\definecolor{m2}{RGB}{250,250,200}
\definecolor{m3}{RGB}{200,250,200}
\begin {figure}
\centering
\resizebox{.45\textwidth}{!}{%
\begin{tikzpicture}[scale=0.75]

    \draw[dashed, fill=m1, draw=black] (1.8, 1.5) rectangle (3.2,-1.5);
    \node[] (min1) at (2.5, 1.0){$\displaystyle\min_{x^1\in\mc X_1}$};
    \node[] (minN) at (2.5, -1.0){$\displaystyle\min_{x^N\in\mc X_N}$};
    
    \draw[dashed, fill=m1, draw=black] (3.3, 1.5) rectangle (6.3,-1.5);
    \node[] (j1) at (4.8, 1.1){$\hat{J}^1\left(x^1,x^{-1}\right)$};
    \node[] (jN) at (4.8, -0.9){$\hat{J}^N\left(x^N,x^{-N}\right)$};
    
    \node[] (p1) at (6.5, 1.1){$+$};
    \node[] (pN) at (6.5, -0.9){$+$};

    \draw[dashed, fill=m3, draw=black] (6.7, 1.5) rectangle (9.4,-1.5);
    \node[] (price1) at (8, 1.1){$\left(x^1\right)^TS_{1}\pi_t$};
    \node[] (priceN) at (8, -0.9){$\left(x^N\right)^TS_{N}\pi_t$};    
     
    \node (d11) at (2.5, 0.3)[circle,fill,inner sep=0.5pt]{};
    \node (d12) at (2.5, 0.0)[circle,fill,inner sep=0.5pt]{};  
    \node (d13) at (2.5, -0.3)[circle,fill,inner sep=0.5pt]{};

    \node (d21) at (4.8, 0.3)[circle,fill,inner sep=0.5pt]{};
    \node (d22) at (4.8, 0.0)[circle,fill,inner sep=0.5pt]{};  
    \node (d23) at (4.8, -0.3)[circle,fill,inner sep=0.5pt]{};

    \node (d31) at (8, 0.3)[circle,fill,inner sep=0.5pt]{};
    \node (d32) at (8, 0.0)[circle,fill,inner sep=0.5pt]{};  
    \node (d33) at (8, -0.3)[circle,fill,inner sep=0.5pt]{};

    \node[] (info1) at (4.1, -1.8){\scriptsize Unknown or partially known};
    \node[] (info11) at (8.1, -1.8){\scriptsize known};

    \draw[] (1.5, 2) rectangle (9.7,-2);
    \node[] (env) at (3.6,2.3){Environment $\mc M\left(\mc C;\pi_t\right)$};
    
    \draw[->] (-0.0, 0) -- (1.5, 0);
    \node[] (pit) at (0.75, 0.3){$\pi_t$};
    \draw[] (-0.5, 0) -- (0, 0);

    \draw[->] (-1.5, 1.5) -- (-1, 1.5);
    \draw[->] (-1.5, -1.5) -- (-1, -1.5);
    
    \draw[] (-1.5, -1) -- (-1.5, -1.5);
    \draw[] (-1.5, -2) -- (-1.5, -1.5);

    \draw[] (-2, -1) -- (-1.5, -1);
    \draw[] (-2, -2) -- (-1.5, -2);    
    
    \node[] (mu) at (-3, -1){$\mu\left(s_t;\theta^{\mu}_t\right)$};
    \node[] (sigma) at (-3, -2){$\Sigma\left(s_t;\theta^{\Sigma}_t\right)$};
    \draw[->] (-4.5, -1) -- (-4, -1);
    \draw[->] (-4.5, -2) -- (-4, -2);

    \draw[] (-6.5, -0.5) rectangle (-4.5,-2.5);
    \node[text width=2cm, text centered] (nn) at (-5.5,-1.5){Neural network};
    
    \draw[] (-4.7, 2) rectangle (-1.5,1);
    \node[text width=2.5cm, text centered] (bf) at (-3.1,1.5){Initial exploration of $\mc P$};

    \draw[->] (-5.5, 1.5) -- (-5.5, -0.5);
    \draw[->] (-5.5, 1.5) -- (-4.7, 1.5);

    \draw[->] (-7.5, -1.5) -- (-6.5, -1.5);
    \node[] (Dt) at (-7,-1.2){$\mc D_t$};
    \draw[] (-9.5, -1) rectangle (-7.5,-2);
    \node[text width=2cm, text centered] (buff) at (-8.5,-1.5){Buffer};
    \draw[->] (-8.5, 2) -- (-8.5, -1);
    \draw[] (-8.5, 2) -- (-5.5, 2);
    \draw[] (-5.5, 2) -- (-5.5, 1.5);

    \node (myfirstpic) at (-7,5.5) {\includegraphics[width=.2\textwidth]{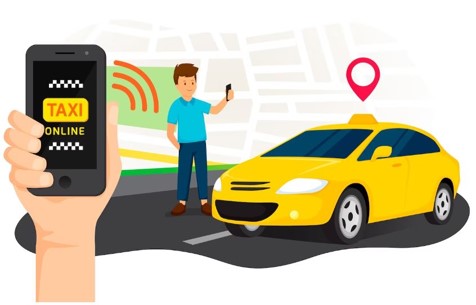}};
    \draw[] (-9.5, 7) rectangle (-4.5,4);
    \draw[->] (-7, 4) -- (-7, 2);
    \draw[] (-11, 3) rectangle (0.0,-3);
    \node[] (Central) at (-9.3,3.3){Central authority $L$};

    \draw[] (-4.5, 5.5) -- (6.5, 5.5);
    \draw[->] (6.5, 5.5) -- (6.5, 2);
    \node[text width=2.6cm] (rhm) at (-7.9,7.6){Ride-hailing demand simulator};

    \node[] (s1) at (-6.7,3.5){$s_t$};
    \node[] (s2) at (-8.2,0.0){$s_t$};
    \node[] (s3) at (-5.2,0.0){$s_t$};
    \node[] (s4) at (6.8,2.5){$s_t$};

    \draw[] (6.5, -2) -- (6.5, -4);
    \draw[] (6.5, -4) -- (-8.5, -4);
    \draw[->] (-8.5, -4) -- (-8.5, -2);
    \node[] (rt) at (-1,-3.6){$R_t\left(\pi_t|s_t\right)= 1-\frac{1}{\sqrt{2}}\norm{\mc Z - \hat{x}^*}_2$};

    \node[] (rt2) at (-8.2,-2.5){$R_t$};
    \draw[-latex] (-0.5,-0.5) arc
    [
        start angle=270,
        end angle=90,
        x radius=0.5cm,
        y radius =0.5cm
    ] ;

    \node (d41) at (-0.5, -0.0)[circle,fill,inner sep=0.75pt]{};
    \draw[] (-0.5, -0.0) -- (-0.95, -1.45);
    
\end{tikzpicture}}%
    \caption{Schematic sketch of the contextual bandit learning setup.} 
\label{fig:problem}
\end{figure}
\section{Electric ride-hailing market}\label{sec:model}
Having defined the optimization problem of every agent in the system, we now proceed to define the elements of the ride-hailing company's cost in more detail. Every company operator is interested in minimizing the one-step-ahead cost under the feasibility constraints imposed by the battery status of its vehicles. Inspired by the objective functions analyzed in~\cite{9838005, Hierarchical, 9102356, article4, 9541309}, in this paper we analyze a cost $J^{i}\left(x^{i},x^{-i};\pi\right) = J_{1}^{i}\left(x^{i}, x^{-i}\right)+J_{2}^{i}\left(x^{i}\right)+J_{3}^{i}\left(x^{i}, \pi\right)$,
where $J_{1}^{i}\left(x^{i}, x^{-i}\right)$ denotes the expected queuing cost, $J_{2}^{i}\left(x^{i}\right)$ is the negative expected revenue and $J_{3}^{i}\left(x^{i}, \pi\right) = \left(x^i\right)^TS_i\pi$ is the charging cost introduced in~\eqref{eq:pricingcost}. 

\textbf{The expected queuing cost} model of the company $i\in\mc C$ effectively depends on the total vehicle distribution as $J_{1}^{i}\left(x^{i}, x^{-i}\right)=\left(x^{i}\right)^{T}C\left(x^i+\overline{\Lambda}x^{-i}-\tau\right)$, 
where $C\in\R^{M\times M}$ is a scaling matrix, i.e., $C=\text{diag}\left(c^i\right)\succeq0$, such that the element $c^i_j\in\R_+$ depicts how expensive it is for a vehicle to queue in the region around the station $j\in\mc H$ and $\tau\in\R^M$ is the vector of charging station capacities. The more the capacity of the station is exceeded, the higher the cost per vehicle should be. Hence, to calculate the total queuing cost for the whole fleet, we take the inner product between the vector describing the fleet's distribution $x^i$, and the incurred cost per vehicle for choosing a particular station.

\textbf{The negative expected revenue} is modeled as $J_{2}^{i}\left(x^{i}\right)=\left(e_{i}^{\text{arr}}\right)^{T}x^{i}-\left(e_{i}^{\text{pro}}\right)^{T}x^{i}$,
where $e_{i}^{\text{arr}}\in \mathbb{R}^{ M}$ is the average cost of a vehicle being unoccupied while traveling to a station and the vector $e_{i}^{\text{pro}}\in \mathbb{R}^{M}$ is the expected profit in regions around charging stations estimated from historical data. Therefore, the part $\hat{J}^i\left(x^i, x^{-i}\right)$ of the total cost in~\eqref{eq:pricingcost} can be simplified to a quadratic form given by
\begin{equation}
    \hat{J}^i\left(x^i, x^{-i}\right)=\frac{1}{2} \left(x^{i}\right)^T P_{i}x^{i}+\left(x^{i}\right)^TQ_i x^{-i}+r_{i}^{T}x^{i} \,,
\label{eq:Ji}
\end{equation}
where $P_i\defineas 2C$, $Q_i\defineas C\overline{\Lambda}$ and $r_i\defineas e_{i}^{\text{arr}}-e_{i}^{\text{pro}}$. For such a game structure, the following proposition guarantees the existence and uniqueness of a Nash equilibrium.
\begin{proposition}\label{prop:1}
    For any $\pi\in\mc P$, let the $\pi-$parametrized ride-hailing market $\mc M\left(\mc C;\pi\right)$ be defined as in~\eqref{eq:market}. Moreover, for every $i\in\mc C$ let the constraint sets $\mc X_i$ be defined as in~\eqref{eq:constrset} and the company operator's objective be defined by~\eqref{eq:Ji}. Under Assumption~\ref{ass:1}, there is a unique v-NE joint strategy $x^*\in\mc X$ describing the interactions between the ride-hailing companies in the market $\mc M\left(\mc C;\pi\right)$. 
\end{proposition}
\iftoggle{full_version}{
\medskip
\noindent The proof of Proposition~\ref{prop:1} is given in Appendix~\ref{app:proof1}.
}{
\medskip
\noindent The proof is given in the extended version of our paper.}\iftoggle{full_version}{
\begin{remark}
\medskip
Under the Assumption~\ref{ass:1}, if the constraint sets are defined by~\eqref{eq:constrset} and the cost function by~\eqref{eq:Ji}, then, for every $i\in\mc C$, the Nash equilibrium strategy $x^{i*}\in\mc X_i$ is the solution of the best-response optimization problem
given by the market definition~\eqref{eq:market}. The optimality of $x^{i*}$ is guaranteed if and only if $x^{i*}$ solves the KKT system of equations
\begin{equation}
\label{eq:KKT}
\left\{\begin{array}{l}
P_ix^{i*}+Q_ix^{-i*}+r_i+S_i\pi+G_i^{\top} \lambda_i^*+\textbf{1}_M \nu_i^*=\mathbf{0}_M \\
\operatorname{diag}\left(\lambda_i^*\right)\left(G_i x^{i*}-h_i\right)=\mathbf{0}_M \\
\textbf{1}_M^T x^{i*}=N_i
\end{array}\right.,
\end{equation}
with $\lambda_i^*\in\R_+^{m_i^{\text{ineq}}}$, $\nu_i^*\in\R$ being the optimal dual variables associated with the inequality and equality constraints. 
\end{remark}
\medskip}{}
\medskip

Given the aggregative structure of the game, several pricing mechanisms and computational methods to find the Nash and local Stackelberg equilibria have already been analyzed in the literature~\cite{9838005,Hierarchical,ECC2023}. However, the underlying assumption in these works entails that all the agents are motivated to work toward the societal optimal described by the central authority's objective. Both the central authority and the ride-hailing companies work together to iteratively compute the local Stackelberg equilibrium. In this paper, we focus on the pricing problem just from the perspective of the central authority. Namely, we aim to investigate what happens if the central authority has no or partial access to $\hat{J}^i\left(x^i, x^{-i}\right)$ and~$\mc X_i$ and hence has to treat the market dynamics as a black-box that outputs the attained v-NE for a particular pricing vector. Therefore, we turn to learning-based methods that involve interacting with the unknown environment and proceed to introduce the framework in the following section.
\section{Learning the charging prices}\label{sec:alg}
At a particular time step $t$, let us assume that the central authority has information about the average charging demand per vehicle and the negative expected revenue, encoded by the vectors $s^d_t\defineas\text{col}\left((d^i)_{i\in\mc C}\right)$ and $s^r_t\defineas\text{col}\left((r^i)_{i\in\mc C}\right)$. Then, the observed state vector of the central authority can be described by $s_t=\text{col}\left(\{s^d_t,s^r_t\}\right)\in\mc K\subseteq\R^{2NM}$. Typical for learning-based optimization problems, in order to encourage exploration of the space of prices, the central authority chooses a pricing vector $\pi_t\in\mc P\subseteq\R^M$ based on the probabilistic pricing policy $\rho\left(\pi|s\right)$. Because the central authority is only focused on optimizing the one-step-ahead cost $J^L\left(x^*(\pi_t),\pi_t\right)$, and there is no clear Markov Decision Process governing the relation between the states $s_t$ and $s_{t+1}$, the optimization problem of the central authority falls under the category of contextual bandits. Hence, we will now present a framework that assumes no prior knowledge about the structure of the electric ride-hailing market.
\subsection{Contextual bandit framework}\label{subsec:bandit}

To cast the central authority's learning problem in the standard form of the contextual bandits, we propose a parametrized form of a Gaussian pricing policy given by
\begin{equation}
    \rho_{\theta}\left(\pi|s\right)\defineas\mc N\left(\mu\left(s;\theta^{\mu}\right), \Sigma\left(s;\theta^{\Sigma}\right)\right)\;.
    \label{eq:rho}
\end{equation}

With this in mind, the central authority now aims to iteratively update $N_{\theta}\in\Z_{>0}$ trainable parameters $\theta=\left[\theta^{\mu}, \theta^{\Sigma}\right]\in\Omega\subseteq\R^{N_{\theta}}$ through interactions with the ride-hailing market in an attempt to maximize the instantaneous reward $R_t\left(\pi|s\right)\in[0,1]$ that describes the quality of the chosen prices. For a particular choice of $\theta$, the objective of the central authority is to maximize the objective $J(\theta)\defineas \mathbb{E}_{\rho_{\theta}\left(\pi|s\right)}\left[R\left(\pi|s\right)\right]$
via policy gradient method first introduced in~\cite{Reinforce}. Namely, to update the parameters of the stochastic policy $\rho_{\theta}\left(\pi|s\right)$ via gradient descent, we aim to utilize a well-known identity concerning $\nabla_{\theta}J(\theta)$ and given by
\begin{equation}
    \label{eq:ddpg} \nabla_{\theta}\mathbb{E}_{\rho_{\theta}\left(\pi|s\right)}\left[R\left(\pi|s\right)\right]=\mathbb{E}_{\rho_{\theta}\left(\pi|s\right)}\left[R\left(\pi|s\right)\nabla_{\theta}\log\rho_{\theta}\left(\pi|s\right)\right]\,.
\end{equation}

The right-hand side of~\eqref{eq:ddpg} is particularly useful if we assume that at time $t=T$, the central authority has access to the history buffer $\mc D_t$ of observed interactions described by triplets $z_t\defineas(s_t, \pi_t, R_t(\pi_t|s_t))$, i.e., $\mc D_t\defineas\left\{z_t\right\}_{t\leq T}$. In that case, the gradient of the central authority's objective, $\nabla_{\theta}J(\theta)=\nabla_{\theta}\mathbb{E}_{\rho_{\theta}\left(\pi|s\right)}\left[R\left(\pi|s\right)\right]$, can be estimated by approximating the expectation on the right-hand side of~\eqref{eq:ddpg} via sampling from $\mc D_t$. To match the objective of the leader introduced in~\eqref{eq:SG}, we set the reward function \iftoggle{full_version}{
\begin{equation}\label{eq:r}
    R\left(\pi|s\right)\defineas 1-\frac{1}{\sqrt{2}}\norm{\mc Z - \hat{x}^*}_2\,,
\end{equation}}
{$R\left(\pi|s\right)\defineas 1-\frac{1}{\sqrt{2}}\norm{\mc Z - \hat{x}^*}_2\in[0,1]$}
such that $x^{*}\in\mc V_{\pi}\left(\mc M\right)$ and $\hat{x}^*=\Lambda x^*/\textbf{1}^Tn$.
\iftoggle{full_version}{
\begin{remark}
    Based on Proposition~\ref{prop:1}, the reward function $R:\mc P\rightarrow[0,1]$ given by~\eqref{eq:r} is well defined. Namely, for every $\pi\in\mc P$, there exists unique $\hat{x}^*$ with $\norm{\hat{x}^*}_{1}=1$. To prove that  $R\left(\pi|s\right)\in[0,1]$, it suffices to observe that
    \begin{equation}\label{eq:rm2}
    \small
    \begin{split}
        \norm{\mc Z-\hat{x}^*}^2_2&=\norm{\mc Z}_2^2+\norm{\hat{x}^*}_2^2-2\mc Z^T\hat{x}^*\leq\norm{\mc Z}_2^2+\norm{\hat{x}^*}_2^2\leq \\
        &\leq\norm{\mc Z}_1^2+\norm{\hat{x}^*}_1^2=1+1=2\,,
    \end{split}
    \end{equation}
    \normalsize
    since the $j$th elements of the vectors $\mc Z$ and $\hat{x}^*$, i.e., $\mc Z_j$ and $\hat{x}^*_j$, satisfy $\mc Z_j,\hat{x}^*_j\in[0,1]$ by construction. 
\end{remark}
\medskip
}{}
By combining~\eqref{eq:rho} and~\eqref{eq:ddpg} at time step $t$, the central authority updates the parameters $\theta$ according to
\small
\begin{equation}
    \argmax_{\theta\in\Omega}\sum_{z_k\in\mc D_t}R_k(\cdot)\sum_{j\in\mc H}\left(\log\frac{1}{\sigma_j(\cdot;\theta)}-\frac{(\pi_{k,j}-\mu_j(\cdot;\theta))^2}{2\sigma_j^2(\cdot;\theta)}\right)\,,
\end{equation}
\normalsize
where $\sigma_j(\cdot;\theta)\in\R_+$ represents the $j$th diagonal element of $\Sigma(s;\theta^{\Sigma})$, $\pi_{k,j}$ represents the $j$th element of the central authority's pricing vector and $\mu_j(\cdot;\theta)\in\R_+$ is the $j$th element of the mean vector $\mu(s;\theta^{\mu})$. 

The complete schematic overview of the learning setup is presented in Figure~\ref{fig:problem}. The ride-hailing EVs serve demand based on the real taxi data from the city of Shenzhen~\cite{BEOJONE2021102890} and the ride-hailing simulator then provides the state of the fleets $s_t$ as an exogenous input to the central authority and the environment. The green part of the environment encapsulates the pricing part of the market that is inherently known to the central authority. On the other hand, the knowledge about the red part depends directly on the willingness of the companies to share information about their operational management. To start the training procedure, the buffer of the observed triplets needs to be filled with some historical data. Needless to say, the quality of the learned pricing policy depends directly on the quality of the observed data. In this paper, we obtain the historical data through initial random exploration of the pricing space $\mc P$. Generally speaking, if there is no prior knowledge about the structure of the model, one would have to explore the space $\mc P=\R^M$ as much as possible. However, for the structure of the market described in Section~\ref{sec:model}, we will propose a method to construct a bounded search space based on the structure of $\hat{J}^i\left(x^i,x^{-1}\right)$ that guarantees attainability of all the interior v-NE $x^*$, i.e., v-NE satisfying $G_ix^{i*}<h_i$, that would be induced by exploring $\R^M$.
\subsection{Characterizing the exploration space}
For a set of pricing policies $\mc P\subseteq\R^{M}$, let the set of $\mc P$-induced interior v-NE of the market game $\mc M\left(\mc C;\pi\right)$ be $\mc V_{\mc P}\defineas\left\{ x^*\in\bigcup_{\pi\in\mc P}\mc V_{\pi}\left(\mc M\right) |\: G_ix^{i*}<h_i,\:\forall i\in\mc C\right\}$.
Let $\overline{c}\in\R_{>0}$ denote the largest diagonal element of $C$, $\overline{N}=\max_{i\in\mc C}N_i$, $\underline{N}=\min_{i\in\mc C}N_i$ and $N_{\text{tot}}=\sum_{i\in\mc C}N_i$. According to the definition of $P_i$ in~\eqref{eq:Ji}, we can drop the subscript $i$ and write $P_i\defineas P$ for every $i\in\mc C$. Furthermore, let $\alpha\in\R_{>0}$ be defined via $\alpha^{-1}=\textbf{tr}(P^{-1})$ and let $\Psi\in\R^{M\times M}$ be $\Psi\defineas\mathbb{I}_M-\alpha\textbf{1}_{M}\textbf{1}^T_MP^{-1}$.
For every $i\in\mc C$, let $\overline{r}_i\defineas\Psi r_i$ and $\overline{r}_{i,j}\in\R$ be its $j$th element. Let $\overline{r}_{\text{max}}=\displaystyle\max_{i\in\mc C,j\in\mc H}\overline{r}_{i,j}$, $\overline{r}_{\text{min}}=\displaystyle\min_{i\in\mc C,j\in\mc H}\overline{r}_{i,j}$, $\overline{z}\defineas\left(\overline{c}-\frac{\alpha}{2}\right)N_{\text{tot}}+\left(\overline{c}+\frac{\alpha}{2}\right)\overline{N}$ and $\underline{z}\defineas\frac{\alpha}{2}\left(\underline{N}-N_{\text{tot}}\right)$.
%
Then the following theorem provides a way to
construct a polytopic exploration space $\mc P$.
\begin{theorem}[Relaxed exploration space]\label{th:1}
    Let the market $\mc M\left(\mc C;\pi\right)$ be defined as in~\eqref{eq:market}. Moreover, for every $i\in\mc C$, let the sets $\mc X_i$ be defined as in~\eqref{eq:constrset} and the company operator's objective be defined by~\eqref{eq:Ji}. If $\overline{\mc P}_1=\R^{M}$, then choosing a polytopic set $\overline{\mc P}_2\defineas\left\{\pi\in\R^M\:|\:\gamma\textbf{1}_{MN}\leq G_{\pi}\pi\leq\Gamma\textbf{1}_{MN}\right\}$
    with $\gamma\defineas\alpha\underline{N}-\overline{r}_{\text{max}}-\overline{z}\in\R$, $\Gamma\defineas\alpha\overline{N}-\overline{r}_{\text{min}}-\underline{z}\in\R$ and $G_{\pi}^T\defineas[S_1^T\Psi^T\:|\:\hdots\:|\:S_N^T\Psi^T]$, yields $\mc V_{\overline{\mc P}_1}=\mc V_{\overline{\mc P}_2}$.
\end{theorem}
\iftoggle{full_version}{
\medskip
\noindent The proof of Theorem~\ref{th:1} is given in Appendix~\ref{app:th1}.
}{
\medskip
\noindent The proof is given in the extended version of our paper.}
\medskip

\iftoggle{full_version}{Uniform sampling from a polytopic constraint is, in general, a hard problem related to approximate calculation of its volume~\cite{randomwalk}. However, based on polytopic $\overline{\mc P}_2$, one can construct a box superset $\overline{\mc P}_3\supseteq\overline{\mc P}_2$ using linear programming.}{Uniform sampling from a polytopic constraint is, in general, a hard problem. However, we show in the extended version how $\mc P_2$ can be extended to a box superset.}\iftoggle{full_version}{
\begin{corollary}
    Let polytope $\overline{\mc P}_2$ be defined as in Theorem~\ref{th:1}. Then, the box polytope $\overline{\mc P}_3\defineas\bigtimes_{j\in\mc H}[\underline{p}_j,\overline{p}_j]$ such that 
    \begin{equation}
        \underline{p}_j=\displaystyle\argmin_{\gamma\textbf{1}_{MN}\leq G_{\pi}\pi\leq\Gamma\textbf{1}_{MN}}\pi_j\:\land\:\overline{p}_j=\displaystyle\argmax_{\gamma\textbf{1}_{MN}\leq G_{\pi}\pi\leq\Gamma\textbf{1}_{MN}}\pi_j\,,
\end{equation}
encapsulates $\overline{\mc P}_2$, i.e., $\overline{\mc P}_3\supseteq\overline{\mc P}_2$.
\end{corollary}
\medskip
It is clear now that uniform sampling from $\overline{\mc P}_3$ boils down to uniform sampling from $[\underline{p}_j,\overline{p}_j]$ for every $j\in\mc H$.}{} 
\iftoggle{full_version}{  
\begin {figure}
\centering
\input{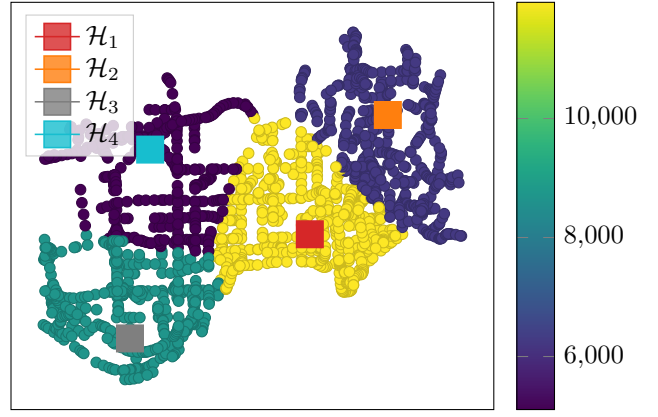}
\caption{The Shenzhen area is divided into four regions around charging stations $\mc H=\left\{\mc H_1,\mc H_2,\mc H_3,\mc H_4\right\}$. The color of the nodes in each region indicates the total number of ride-hailing requests originating in that region. The highest number occurs around $\mc H_1$ and the lowest around $\mc H_4$.}
\label{fig:Map}
\end{figure}}{}
Finally, if the central authority also has knowledge about the inequality constraints of the ride-hailing companies, we can show that the learning process can be completely reduced to an instance of a mixed integer program. 
\subsection{Complete knowledge about the market}\label{sec:completemarket}
If the central authority has full knowledge about the cost functions and constraint sets in $\mc M(\mc C)$, then the complete learning procedure can be avoided by solving an MPCC.
\begin{theorem}[Global optimum feasibility check]\label{th:2}
    Let the $\pi-$parametrized ride-hailing market $\mc M\left(\mc C;\pi\right)$ be defined as in~\eqref{eq:market}. Moreover, for every $i\in\mc C$, let the constraint sets $\mc X_i$ be defined as in~\eqref{eq:constrset} and the company operator's objective be defined by~\eqref{eq:Ji}. There exists a vector $\pi\in\R^{M}$ such that $J^{L*}\left(\cdot,\pi\right)=0$ if and only if there exists $\beta>0$ such that following feasibility MILP has a solution
    \begin{mini!}
    {x^*, \lambda^*, \nu^*, \pi, \textbf{m}}{1}
    {}{}
    \addConstraint{ \hspace{-0.4cm}\overline{\textbf{P}}_1x^*+\overline{\textbf{P}}_2\nu^*+\overline{\textbf{P}}_3\lambda^*=\overline{\textbf{r}}+\overline{\textbf{S}}\pi}{}{\label{eq:cc1}}
    \addConstraint{\hspace{-0.4cm}\overline{\textbf{A}}x^*=\overline{\textbf{b}}}{}{\label{eq:cc2}}
    \addConstraint{\hspace{-0.4cm}\Lambda x^*=\textbf{1}^T_Mn\mc Z}{}{\label{eq:cc3}}
    \addConstraint{\hspace{-0.4cm}\textbf{0}_L\leq\lambda^*\leq \beta \textbf{m}}{}{\label{eq:cc4}}
    \addConstraint{\hspace{-0.4cm}\textbf{0}_L\leq -\left(\overline{\textbf{G}}x^*-\overline{\textbf{h}}\right)\leq \beta(\textbf{1}_L-\textbf{m})}{}{\label{eq:cc5}}
    \addConstraint{\hspace{-0.4cm}\textbf{m}\in\{0,1\}^L}{}
    \label{mini:op1}
    \end{mini!}
where $L=\sum_{i\in\mc C}m_i^{\text{ineq}}$, $\overline{\textbf{P}}_1=\mathbb{I}_{N}\otimes C+\mathbf{1}_{N}\mathbf{1}_{N}^T\otimes C$, $\overline{\textbf{P}}_2=\text{Diag}(\textbf{1}_M)_{i\in\mc C}$, $\overline{\textbf{P}}_3=\text{Diag}(G_i^T)_{i\in\mc C}$, $\overline{\textbf{r}}=\text{col}((r_i)_{i\in\mc C})$,
$\overline{\textbf{A}}=\text{Diag}(\textbf{1}_M^T)_{i\in\mc C}$,
$\overline{\textbf{b}}=\text{col}((N_i)_{i\in\mc C})$,
$\overline{\textbf{G}}=\text{Diag}(G_i)_{i\in\mc C}$,
$\overline{\textbf{h}}=\text{col}((h_i)_{i\in\mc C})$,
$\overline{\textbf{S}}^T=[S_1\:|\:\hdots\:|\:S_M]$.   
\end{theorem}
\iftoggle{full_version}{
\medskip
\noindent The proof of Theorem~\ref{th:2} is given in Appendix~\ref{app:th2}.
}{
\medskip
\noindent The proof is given in the extended version of our paper.}
\medskip

However, if the exact minimization of the leader's objective $J^L(\cdot,\pi)$ is not possible, i.e., the feasibility MILP in Theorem~\ref{th:2} does not have a solution, one can search for the optimal pricing vector $\pi\in\mc P$ by solving a MIQP obtained by removing the constraint~\eqref{eq:cc3} and changing the objective in the feasibility MILP to $\lVert\Lambda x^*-\textbf{1}^Tn\mc Z\lVert_2^2$. Generally speaking, in the case of full knowledge about the structure of the market $\mc M(\mc C)$, avoiding the learning process is done at the expense of having to find a proper parameter $\beta>0$ for the big-M reformulation. This is an NP-hard problem~\cite{RePEc}, tackled in reality via different heuristics~\cite{KLEINERT2021100007}. However, for a dynamic scenario in which the parameters describing the cost functions $J^{i}$ of the ride-hailing companies change in each iteration based on the state of the respective ride-hailing fleet, applying such a method might be less favorable than having a trained pricing agent.

In the next section, we will present the results obtained in a simulated case study based on real taxi data from Shenzhen.
\section{Case study}\label{sec:example}
Let us assume there exist three ride-hailing companies $\mc C=\left\{\mc C_1, \mc C_2, \mc C_3\right\}$ with fleet sizes $N_{\text{fleet}}=\left[450, 400, 350\right]^{T}$ that serve the ride-hailing demand in the Shenzhen region with four public charging stations $\mc H=\left\{\mc H_1, \mc H_2, \mc H_3, \mc H_4\right\}$. The stations are located in parts of Shenzhen with different demands for ride-hailing services\iftoggle{full_version}{ as shown in the color-coded map depicted in Figure \ref{fig:Map}}{}
 and are described by the vector of capacities $\tau=\left[15, 60, 35, 50\right]^{T}$. After a 3-hour long simulation, representing one of the two peak-hour periods during the day when the companies serve the real taxi demand obtained from~\cite{BEOJONE2021102890}, the vehicles whose battery level is lower than 55\% are considered interested in charging. To prevent the ride-hailing vehicles from flocking in the most demand-attractive parts of Shenzhen, the desired distribution of the ride-hailing vehicles $\mc Z$ is formed so as to match the spatial distribution of the ride-hailing service requests. To approximate this distribution, the city is divided into four cells according to the Voronoi~\cite{Kang2008} partitioning of the map, with the stations chosen as the centroids of the Voronoi cells. The distribution $\mc Z$ is chosen to correspond to the total number of requests in each cell which results in $\mc Z=\left[0.37, 0.19, 0.27, 0.17\right]^T$.
\iftoggle{full_version}{\begin {figure}
\centering
\resizebox{.45\textwidth}{!}{%
\input{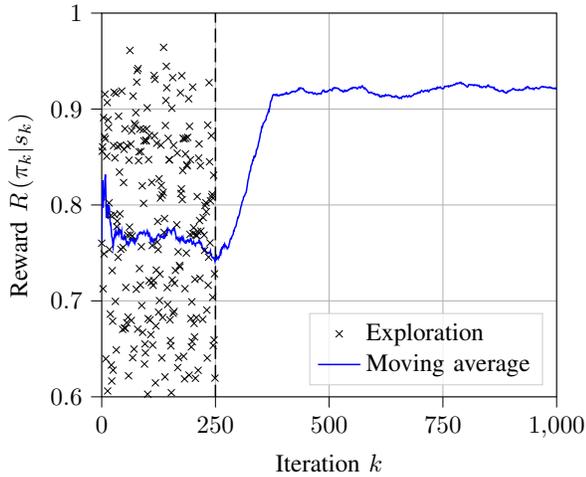}}
\caption{Evolution of the reward signal. The solid blue line represents the moving average of the last 100 iterations.}
\label{fig:reward}
\end{figure}}{\begin {figure}
\centering
\resizebox{.35\textwidth}{!}{%
\input{figures/reward.tikz}}
\caption{Evolution of the reward signal. The solid blue line represents the moving average of the last 100 iterations.}
\label{fig:reward}
\end{figure}}
 \iftoggle{full_version}{ For the mean $\mu(s;\theta^{\mu})$ and the covariance $\Sigma\left(s;\theta^{\Sigma}\right)$ we use the same neural network structure containing three hidden layers with 256, 64 and 16 nodes respectively. After each hidden layer we use the 'ReLu' activation function, whereas after the output layer we use the 'sigmoid' for the mean neural network and 'softplus' for the covariance neural network~\cite{activation}.}{} We run the contextual bandit for a total of $N_{\text{iter}}=1000$ iterations, with the first $N_{\text{exp}}=250$ used for random exploration. In the $t$-th iteration, the central authority samples a batch $\mc B_t$ of $|\mc B_t|=32$ triplets from the observations in the current buffer $\mc D_t$ and uses them to perform $N_{\text{epoch}}=20$ epochs of the parameter update procedure. After the update has been completed, for the exogenously given state $s_t$, the agent performs the forward propagation to obtain $\mu(s_t;\theta^{\mu}_t)$ and $\Sigma(s_t;\theta^{\Sigma}_t)$, and then samples the pricing policy $\pi_t\sim\mc N(\mu(s_t;\theta^{\mu}_t), \Sigma(s_t;\theta^{\Sigma}_t))$ for the current iteration. The pricing is then applied in the market $\mc M(\mc C)$, the resulting distribution of vehicles $\hat{x}^*$ among the stations is observed, the reward $R_t(\pi_t|s_t)$ is calculated and the resulting triplet is stored.
\iftoggle{full_version}{
\begin {figure}
\centering
\resizebox{.45\textwidth}{!}{%
\input{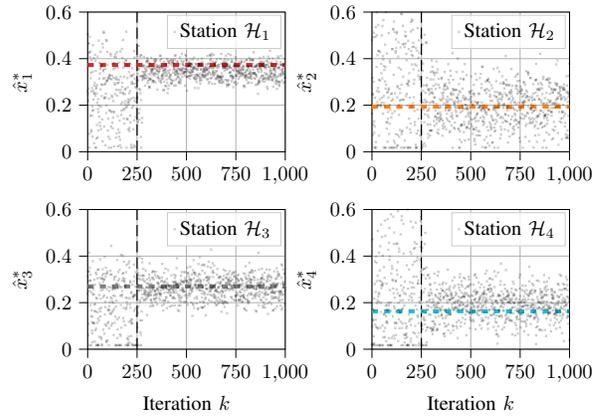}}
\caption{Evolution of the achieved distribution $\hat{x}^*$. In each subplot, the dashed line corresponds to $\mc Z_j$ and shows the desired value of the particular component of the attained distribution $\hat{x}^*$. The grey marks show the attained value of $\hat{x}^*_j$ at each iteration.}%
\label{fig:dist}
\end{figure}}{}
The performance of the contextual bandit is depicted in \iftoggle{full_version}{Figures~\ref{fig:reward} and~\ref{fig:dist}. In Figure~\ref{fig:reward},}{Figure~\ref{fig:reward}, where} the blue line shows how the moving average of 100 samples of the reward signal evolves with respect to the iteration number. The black marks on the left-hand side of the dashed line denoting the 250-th iteration show the attained rewards during the training data collection phase. It is important to note here that the exogenous state of the system $s_k$ is different at every iteration meaning that the system aims to map a whole distribution of ride-hailing market states into reasonably good pricing vectors. When learning begins, we can observe that the moving average curve seemingly converges after approximately 150 iterations to a value that yields good matching between the desired and attained vehicle distribution. \iftoggle{full_version}{This is further supported by Figure~\ref{fig:dist} where we compare the individual components of the desired and attained vehicle distributions, i.e., $\mc Z$ and $\hat{x}^*$. The grey dots represent the exact values of the particular component of $\hat{x}^*$, whereas the dashed line shows the corresponding component value of $\mc Z$. We have to keep in mind that the pricing vector is sampled in every iteration, hence the dispersion of the grey dots in the subplots.}{We further support this by plotting the evolution of attained vehicle distributions in the extended version of the paper.} Regarding the collection of the training data during the initial exploration phase, based on the results from~\cite{ECC2023,Hierarchical}, we a priori chose $\mc P=[0.0,5.0]^4$. Since the exogenously generated state $s_k$ determines the parameters of the cost functions in the market for every iteration, one should verify that the sampled $\pi\in\mc P$ also belongs to the set $\overline{\mc P}_2$ given by Theorem~\ref{th:1}. We observed that for the parameters of our case study, there was a significant gap between $\overline{\mc P}_2$ and $\mc P=[0.0,5.0]^4$, which stresses the fact that the bounds derived in Theorem~\ref{th:1} are not tight. In the future, we aim to investigate how and if these bounds can be improved. Finally, for a particular state $s$ for which the feasibility MILP in Theorem~\ref{th:2} was feasible, we compare the performance of the pricing policy obtained by the feasibility MILP and the fully trained contextual bandit. The comparison of the charging prices and attained vehicle distributions is displayed in Table~\ref{tab:res}.  
\begin{table}
\begin{center}
 \renewcommand{\arraystretch}{1.2}
 \caption{Charging prices and attained vehicle distributions}\vspace{1ex}
 \label{tab:res}
 \begin{tabular}{c|c|cc|cc}
 \multirow{2}{*}{Station $\mc H$} & \multirow{2}{*}{$\mc Z$} & \multicolumn{2}{c|}{Contextual Bandit} & \multicolumn{2}{c}{Feasibility MILP}  \\ 
 & &$\pi_{j}$ & $\hat{x}^*_j$ &$\pi_{j}$ & $\hat{x}^*_j$ \\
 \hline
 \hline
 $\mc H_{1} \rule{0pt}{2.6ex}$ &  0.37 & 3.69 & 0.35 & 3.39  & 0.37   \\
 $\mc H_{2}$                   &  0.19 & 2.37 & 0.20 & 2.20  & 0.19   \\
 $\mc H_{3}$                   &  0.27 & 2.87 & 0.26 & 2.83  & 0.27   \\
 $\mc H_{4}$                   &  0.17 & 1.34 & 0.19 & 1.58  & 0.17   \\ 
\hline 
\end{tabular}
\end{center}
\end{table}
The illustrated numerical values are obtained for $\pi=\mu(s;\theta^{\mu})$. Clearly, the MILP algorithm achieves $R_{\text{MILP}}=1.0$. The CB-agent, on the other hand, achieves $R_{\text{CB}}=0.974$, which represents a minor discrepancy between the desired and attained distributions. 
\section{Conclusions}\label{sec:conclusion}
In this paper, we present a learning algorithm based on the concept of contextual bandits. It is capable of learning how to map an exogenously given distribution of the ride-hailing fleet states into an adequate charging price vector for a ride-hailing market described by a quadratic aggregative game. For the case when no domain knowledge is available, but the central authority has access to the form of the cost functions in the market, we propose a polytopic search space for generating the training data in the initial phase of the process. If the central authority has full knowledge of the market structure, we show that it suffices to solve an instance of a MILP/MIQP in order to find the optimal pricing. 

In the future, we aim to investigate if the bounds on the search space in Theorem~\ref{th:1} can be tightened and if more complex instances of the pricing games could be solved.

\bibliographystyle{IEEEtran}
\bibliography{references.bib}

\iftoggle{full_version}{
\appendix
\subsection{Proof of Proposition~\ref{prop:1}}\label{app:proof1}
\begin{proof}
    Since $P_i\succ 0$ for every $i\in\mc C$, the agents' cost functions are convex in $x^i$. For $\mc X_i$ defined as in~\eqref{eq:constrset}, based on~\cite[T.1]{Rosen}, there exists a Nash equilibrium of the $\pi$-parametrized market game. A sufficient condition for the uniqueness of the Nash equilibrium is that the operator $F\left(x,\pi\right)$ be strictly monotone in $x$~\cite[Ch.2]{VIproblems}. Based on~\eqref{eq:Ji}, the pseudo gradient can be written as $F\left(x,\pi\right)=F_1x+F_2$, such that $F_1=\mathbb{I}_{N}\otimes C+\mathbf{1}_{N}\mathbf{1}_{N}^T\otimes C$ and 
    $F_2=\text{col}\left(\left(r_i+S_i\pi\right)_{i\in\mc C}\right)$. To show that $F\left(x,\pi\right)$ is strictly monotone, it suffices to prove that $F_1\succ 0$~\cite{ConvexAnalysis}. This is true as for any $x\in\mc X$, it holds that
    $x^TF_1x=\sum_{i\in\mc C}\left(x^i\right)^TCx^i+\left(\sum_{i\in\mc C}x^i\right)^TC\left(\sum_{i\in\mc C}x^i\right)>0$.
\end{proof}

\subsection{Proof of Theorem~\ref{th:1}}\label{app:th1}
\begin{proof}
Let $\pi\in\R^M$ be any pricing policy that yields a v-NE $x^*\in\mc X$ such that for all $ i\in\mc C$ it holds that $G_ix^{i*}<h_i$. We will show that $\pi\in\overline{\mc P}_2$ as well. Based on~\eqref{eq:Ji}, for every $i\in\mc C$, the KKT conditions~\eqref{eq:KKT} boil down to
\begin{equation}\label{eq:matkkt}
    \left[\begin{array}{cc}
         P & \textbf{1}_M \\
         \textbf{1}_M^T & 0 
    \end{array}\right]\left[\begin{array}{c}
         x^{i*}  \\
         \nu_i^{*} 
    \end{array}\right]=\left[\begin{array}{c}
         -\left(C\Lambda_{-i}x^*+r_i+S_i\pi\right)  \\
         N_i 
    \end{array}\right]\,,
\end{equation}
since $G_ix^{i*}<h_i$ yields $\lambda^{*}_i=\textbf{0}$. Using the Matrix Inversion Lemma~\cite{Matrix} and noting that $(\textbf{1}_M^TP^{-1}\textbf{1}_M)^{-1}=\alpha$, we obtain
\begin{equation}\label{eq:nusol}
    \nu_i^{*}=-\alpha\textbf{1}^T_MP^{-1}\left(C\Lambda_{-i}x^*+r_i+S_i\pi\right)\,.
\end{equation}
For every $i\in\mc C$, let us define $z_i\defineas\left(P\Lambda_i+\Psi C\Lambda_{-i}\right)x^*$. Then, by inserting~\eqref{eq:nusol} into~\eqref{eq:matkkt}, we get
\begin{equation}\label{eq:psipi}
    \Psi S_i\pi=\alpha N_i\textbf{1}_M - \Psi r_i-z_i\,.
\end{equation}
Observe that we can transform $z_i$ into $z_i=\Psi C\sum_{j\in\mc C}x^{j*}+(\mathbb{I}_M+\alpha\textbf{1}_{M}\textbf{1}^T_MP^{-1})Cx^{i*}$. Note that $\Psi C=C-\frac{\alpha}{2}\textbf{1}_M\textbf{1}^T_M$ and $(\mathbb{I}_M+\alpha\textbf{1}_{M}\textbf{1}^T_MP^{-1})C=C+\frac{\alpha}{2}\textbf{1}_M\textbf{1}^T_M$ because $P=2C$. Based on the Harmonic Mean-Arithmetic Mean (HM-AM) inequality, we have that
\begin{equation}
    M\alpha=\frac{M}{\sum_{j\in\mc H}P^{-1}_{j,j}}\leq\frac{\sum_{j\in\mc H}P_{j,j}}{M}\leq\frac{M\cdot2\overline{c}}{M}=2\overline{c}\,.
\end{equation}
Let $z_{i,j}\in\R$ denote the $j$th element of the vector $z_i$. Given that $\overline{c}-\frac{\alpha}{2}\geq\overline{c}-\frac{\overline{c}}{M}>0$, we can write $\max_{j\in\mc H}z_{i,j}\leq\left(\overline{c}-\frac{\alpha}{2}\right)\sum_{i\in\mc C}N_i+\left(\overline{c}+\frac{\alpha}{2}\right)N_i\leq \overline{z}$. Similarly, we have that $\min_{j\in\mc H}z_{i,j}\geq-\frac{\alpha}{2}\sum_{i\in\mc C}N_i+\frac{\alpha}{2}N_i\geq\underline{z}$. Combining this with~\eqref{eq:psipi}, we have that for every $i\in\mc C$ it holds that
\begin{equation}\label{eq:inequality}
\begin{split}
       &\Psi S_i\pi\leq\alpha\overline{N}\textbf{1}_M-\overline{r}_{\text{min}}\textbf{1}_M-\underline{z}\textbf{1}_M \\
       &\Psi S_i\pi\geq\alpha\underline{N}\textbf{1}_M-\overline{r}_{\text{max}}\textbf{1}_M-\overline{z}\textbf{1}_M\,.
\end{split}
\end{equation}
Finally, stacking~\eqref{eq:inequality} for every $i\in\mc C$ shows that $\pi\in\overline{\mc P}_2$.
\end{proof}

\subsection{Proof of Theorem~\ref{th:2}}\label{app:th2}
\begin{proof}
It is clear that the constraint~\eqref{eq:cc3} guarantees that $J^{L*}\left(\cdot,\pi\right)=0$. Based on~\eqref{eq:Ji}, the constraints~\eqref{eq:cc1} and~\eqref{eq:cc2} are obtained by stacking together the first and the third KKT condition in~\eqref{eq:KKT} for all $i\in\mc C$. The constraints~\eqref{eq:cc4} and~\eqref{eq:cc5} represent the big-M reformulation~\cite{bigm} of the stacked complementary slackness constraints in~\eqref{eq:KKT}. Based on~\cite{bigm}, for a properly chosen $\beta>0$, solving the proposed MILP is equivalent to solving the set of KKT best-response optimization problems~\eqref{eq:KKT} for all $i\in\mc C$.  
\end{proof}
}

{}
\end{document}